# The ratio of normal state to superconducting state of spin lattice relaxation rate of unconventional superconductors

P.Udomsamuthirun* and J.Wanghunklang

Department of Physics, Faculty of Science, Srinakharinwirot University, Sukumvit 23 Road, Bangkok 10110,Thailand

**Abstract**

In this research, we derive a simple expression for the ratio of normal state superconducting state of spin lattice relaxation rate of unconventional superconductors from the BCS weak-coupling equation. The unconventional superconductors we consider have three types of order parameters as d-wave, $3He$ A-phase and p-wave 3-dimensional order parameter that had been done before by Parker and Haas[7]. After using some numerical approximations and some boundary conditions, we can find the ratio of normal state to superconducting state of spin lattice relaxation rate in power series of temperature dependent order parameters and temperature. Our numerical calculations show the coherence peak below critical temperature clearly that are consistent with Parker and Haas[7]. These results do not agree with the believed that the coherence peak is the only property of s-wave superconductor. However from our calculation, we can conclude that the unconventional superconductors can show the coherence peaks.

# 1. Introduction

One of the powerful tools for analyzing ordered electrons in solid is Nuclear magnetic resonance(NMR). In superconductor, it has been used for analyzing the symmetry of the superconducting order parameters[1] and clarify the structure of phase diagram in systems with competing orders[2] . It is well know[3] that conventional superconductors obeying BCS weak-coupling theory generally show a large coherence peak in the nuclear spin lattice relaxation rate $T_1^{-1}$ below $T_c$ and the unconventional superconductors show no coherence peak near $T_c$.

Hasegawa[4] considered the density of state in an anisotropic superconductor with intersecting line nodes in the gap function is proportional to $E \log(\alpha \Delta_0 / E)$ for $|E < \Delta_0|$, where $\Delta_0$ is the maximum value of the gap function and α is constant, while it is proportional to E if the line nodes do not intersect. He found that a logarithmic correction appear in temperature dependence of the NMR relaxation rate. He also calculated the NMR relaxation rate $T_1^{-1}$ by assuming the temperature dependence of the maximum gap to be $\Delta_0(T) = 1.76 T_c \tanh(1.74 \sqrt{\frac{T_c}{T} - 1})$ which is the approximate temperature dependence of the energy gap in s-wave superconductors in the weak coupling limit. Hayashi and et al.[5] studied numerically the nuclear spin lattice relaxation $T_1^{-1}$ around a single vortex in a noncentrosymmetric superconductor for s+p-wave Cooper pairing. Rostunov, Demler and Georges[6] studied spin-wave excitations in triplet superconductors. They introduced an effective bosonic model to describe spin-wave excitations and calculated their contribution to the nuclear spin lattice relaxation rate. They showed that $T_1^{-1}$ has a power law scaling with temperature, including the $T^3$ and $T^5$ dependence for two-dimensional and three-dimensional systems, respectively. Parker and Haas [7] showed that an unconventional superconductors ; d-wave, $^3$He A-phase and 3D-linenode; obeying BCS pure-case weak-coupling theory should show a small $T_1^{-1}$ coherence peak near $T_c$. For three cases of consideration, they use the approximation that near $T_c$ the density of state is $N(E) = F(E) +1$ and $F(E)$ only varies from zero and derived an analytic expression for density of state and $T_1^{-1}$.

In this work, we review the works of Parker and Haas[7] that consider the three cases of unconventional superconductors ; d-wave, $^3$He A-phase and p-wave 3-dimensional order parameter to derive an analytic expression for the ratio of normal state to superconducting state of spin lattice relaxation rate . We calculate the ratio of normal state to superconducting state of spin lattice relaxation rate from the density of state $N(\varepsilon)$ and compute numerically by using temperature dependent order parameters for different of $T_c$ and $\omega_D$.

# 2. Model and Calculation

Within the BCS weak-coupling equation, the ratio of the nuclear spin lattice relaxation rate of normal state to superconducting state for unconventional superconductors is defined as[4,7]

$$\frac{(T_1 T)^{-1}}{(T_1 T)^{-1}\big|_{T=T_c}} = \int_0^{\omega_D} d\varepsilon N^2(\varepsilon) \frac{\sec h^2(\frac{\varepsilon}{2T})}{2T} \tag{1}$$

Where $\frac{1}{T_1}$ and $\frac{1}{T_1}\big|_{T=T_c}$ is the nuclear spin lattice relaxation rate of superconducting state and normal state respectively. $N(\varepsilon)$ is the density of state calculated from BCS expression. $\omega_D$ is Debye cut off energy.

Here we consider the ratio of the nuclear spin lattice relaxation rate of normal state to superconducting state near $T_c$ for 3 type of order parameters such as $d_{x^2-y^2}$-wave, the 3He A-phase, p-wave 3-dimensional order parameter. The order parameters are defined as

$$\Delta_k(T) = \Delta(T) f(k) \tag{2}$$

Here $\Delta_k(T)$ is temperature and angular dependent order parameters. $\Delta(T)$ is the temperature dependent order parameter. And $f(k)$ is an angular dependent function. The gap $\Delta(T)$ can be found by using the BCS gap equation. For more accuracy, the same process as [8-9] is used to find $\Delta(T)$ with determined $\Delta(0)$, $T_c$ and $\omega_D$. The temperature dependent energy gap can get from the equation

$$\int_{-\omega_D}^{\omega_D} d\varepsilon \frac{\tanh(\varepsilon/2T_c)}{\varepsilon} = \frac{1}{2\pi} \int_0^{2\pi} d\theta \int_{-\omega_D}^{\omega_D} d\varepsilon \frac{N(\varepsilon, \Delta(T)) \tanh(\sqrt{\varepsilon^2 + \Delta_k^2(T)}/2T)}{\sqrt{\varepsilon^2 + \Delta_k^2(T)}} \tag{3}$$

with zero temperature gap $\Delta(0)$ as

$$\int_{-\omega_D}^{\omega_D} d\varepsilon \frac{\tanh(\varepsilon/2T_c)}{\varepsilon} = \frac{1}{2\pi} \int_0^{2\pi} d\theta \int_{-\omega_D}^{\omega_D} d\varepsilon \frac{N(\varepsilon, \Delta(0))}{\sqrt{\varepsilon^2 + \Delta_k^2(0)}} \tag{4}$$

Here $\Delta(0)$ and $\Delta(T)$ are the order parameter at zero temperature and at temperature $T$, respectively.

Case I. $d_{x^2-y^2}$-wave superconductors

As we have known before that high temperature superconductors such as $Y123$-superconductor is the $d_{x^2-y^2}$-wave superconductors, the order parameter is $\Delta_k^d(T) = \Delta^d(T) \sin(2\theta)$ or $\Delta_k^d(T) = \Delta^d(T) \cos(2\theta)$ in 2-dimensional system. So in this case, we can get $f(k) \equiv f(\theta) = \cos(2\theta)$ or $f(\theta) = \sin(2\theta)$ and the density of state can be calculated from BCS expression

$$N(\varepsilon) = \text{Re} < \frac{\varepsilon}{\sqrt{\varepsilon^2 - \Delta_k^2(T)}} > \tag{5}$$

where $< ... >$ denotes an average over the Fermi surface.

After some calculations, we can get the density of state as [7,10]

$$N(x) = \frac{2}{\pi} \kappa(\frac{1}{x}) \qquad , x > 1;$$

$$= \frac{2}{\pi} x\kappa(x) \quad , \quad x < 1 . \tag{6}$$

Here $x = \frac{\varepsilon}{\Delta}$ and $\kappa(x)$ is the complete elliptic integral of the first kind.

Substitution Eq.(6) into Eq.(1), we find

$$\frac{(T_1T)^{-1}}{(T_1T)^{-1}\big|_{T=T_c}} = \frac{\Delta}{2T}\int_0^1 dx(\frac{2}{\pi}x\kappa(x))^2 \operatorname{sech}^2(\frac{\Delta x}{2T}) + \frac{\Delta}{2T}\int_1^{\omega_D/\Delta} dx(\frac{2}{\pi}\kappa(\frac{1}{x}))^2 \operatorname{sech}^2(\frac{\Delta x}{2T}) \tag{7}$$

Because $\operatorname{sech}^2 x = 1 - \tanh^2 x$ and $\tanh x \approx x, x<1$ and $\tanh x \approx 1, x>1$. Eq.(7) must be separated into two case; $\Delta < 2T$ and $\Delta > 2T$.

For $\Delta < 2T$, we get

$$\frac{(T_1T)^{-1}}{(T_1T)^{-1}\big|_{T=T_c}} = \frac{\Delta}{2T}\int_0^1 dx(\frac{2}{\pi}x\kappa(x))^2(1-(\frac{\Delta x}{2T})^2) + \frac{\Delta}{2T}\int_1^{2T/\Delta} dx(\frac{2}{\pi}\kappa(\frac{1}{x}))^2(1-(\frac{\Delta x}{2T})^2) + a$$

Here the higher order term is defined as constant $a$. The numerical calculations are

$$\int_0^1 dx(\frac{2}{\pi}x\kappa(x))^2 = 0.6424, \quad \int_0^1 dx(\frac{2}{\pi}x^2\kappa(x))^2 = 0.4534 \text{ and}$$

$$\frac{\Delta}{2T}\int_1^{2T/\Delta} dx(\frac{2}{\pi}\kappa(\frac{1}{x}))^2 - (\frac{\Delta x}{2T})^3\int_1^{2T/\Delta} dx(\frac{2}{\pi}x\kappa(\frac{1}{x}))^2 \approx 0.6680 - 0.2920\frac{\Delta}{2T}$$

Within above equation, we can get

$$\frac{(T_1T)^{-1}}{(T_1T)^{-1}\big|_{T=T_c}} = (0.6680 + a) + 0.3504\frac{\Delta}{2T} - 0.4534(\frac{\Delta}{2T})^3 \tag{8}$$

For $\Delta > 2T$, we get

$$\frac{(T_1T)^{-1}}{(T_1T)^{-1}\big|_{T=T_c}} = \frac{\Delta}{2T}\int_0^{2T/\Delta} dx(\frac{2}{\pi}x\kappa(x))^2 - (\frac{\Delta}{2T})^3\int_1^{2T/\Delta} dx(\frac{2}{\pi}x^2\kappa(\frac{1}{x}))^2 + b(\frac{2T}{\Delta})^\alpha + c$$

Here the higher order terms are defined as $b(\frac{2T}{\Delta})^\alpha + c$ where b, c, $\alpha$ are constant.

The numerical calculations are

$$\frac{\Delta}{2T}\int_0^{2T/\Delta} dx(\frac{2}{\pi}x\kappa(x))^2 \approx \frac{(2T/\Delta)^2}{3} + \frac{(2T/\Delta)^4}{10}$$

$$(\frac{\Delta}{2T})^3\int_0^{2T/\Delta} dx(\frac{2}{\pi}x^2\kappa(x))^2 \approx \frac{(2T/\Delta)^2}{5} + \frac{(2T/\Delta)^4}{14} .$$

We can get

$$\frac{(T_1T)^{-1}}{(T_1T)^{-1}\big|_{T=T_c}} = 0.1333(\frac{2T}{\Delta})^2 + 0.0286(\frac{2T}{\Delta})^4 + b(\frac{2T}{\Delta})^\alpha + c \qquad (9)$$

There are 3 conditions for Eq.(8) and Eq.(9); 1) At $T=T_c$, $\frac{(T_1T)^{-1}}{(T_1T)^{-1}\big|_{T=T_c}} = 1$. 2) At $T=0$, $\frac{(T_1T)^{-1}}{(T_1T)^{-1}\big|_{T=T_c}} =$ constant. However, for simplicity, let $\frac{(T_1T)^{-1}}{(T_1T)^{-1}\big|_{T=T_c}} = 0$

And 3) It must be continuous function with continuous first derivative at $\Delta = 2T$.

After taking above conditions, the $\frac{(T_1T)^{-1}}{(T_1T)^{-1}\big|_{T=T_c}}$ in two regions of consideration becomes

$$\frac{(T_1T)^{-1}}{(T_1T)^{-1}\big|_{T=T_c}} = 1 + 0.3504\frac{\Delta(T)}{2T} - 0.4534(\frac{\Delta(T)}{2T})^3 \qquad \text{,for } \Delta(T) < 2T$$

$$= 0.1333(\frac{2T}{\Delta(T)})^2 + 0.0286(\frac{2T}{\Delta(T)})^4 + 0.7351(\frac{2T}{\Delta(T)})^{0.86} \text{,for } \Delta(T) > 2T$$

(10)

Eq.(10) is the simplified equation of the ratio of the nuclear spin lattice relaxation rate of superconducting state to normal state for $d_{x^2-y^2}$-wave superconductors that shown in the power series of $\Delta(T)$. Using Eqs.(3-4) for calculating $\Delta(T)$ and substitution into Eq.(10), the ratio of the nuclear spin lattice relaxation rate of superconducting state to normal state for $d_{x^2-y^2}$-wave superconductors is shown in Figure.1. It is clear that there is the coherence peak near $d_{x^2-y^2}$-wave superconductors.

Case II. $^3$He A-phase

$^3$He A-phase is an example of the condensed matter systems that show the superfluidity of Fermion system. This superfluidity has the same mechanism as superconductors so the measurement of the nuclear spin-lattice relaxation should also be the same as superconductors. We can consider it as one of the superconductor's example. The order parameter of $^3$He A-phase is $\Delta_k^{He}(T) = \Delta^{He}(T)\sin\theta$ in 3–dimensional system. In this case, we can get $f(k) \equiv f(\theta) = \sin\theta$. The density of state can be calculated from Eq.(5), that [7]

$$N(x) = \frac{x}{2}\ln\left|\frac{1+x}{1-x}\right| \qquad (11)$$

Here $x = \frac{\varepsilon}{\Delta}$. Substitution Eq.(11) into Eq.(1), we find

$$\frac{(T_1T)^{-1}}{(T_1T)^{-1}\big|_{T=T_c}} = \frac{\Delta}{2T}\int_1^{\omega_D/\Delta} dx(\frac{x}{2}\ln\left|\frac{1+x}{1-x}\right|)^2 \operatorname{sech}^2(\frac{\Delta x}{2T}) \qquad (12)$$

Within the same process, we get for $\Delta < 2T$ as

$$\frac{(T_1T)^{-1}}{(T_1T)^{-1}\big|_{T=T_c}} = \frac{\Delta}{2T}\int_0^1 dx(\frac{x}{2}\ln(\frac{1+x}{1-x}))^2(1-(\frac{\Delta x}{2T})^2) + \frac{\Delta}{2T}\int_1^{2T/\Delta} dx(\frac{x}{2}\ln(\frac{1+x}{x-1}))^2(1-(\frac{\Delta x}{2T})^2) + d$$

$$= (0.6690 + d) + 0.6426\frac{\Delta}{2T} - 0.4975(\frac{\Delta}{2T})^3$$

Here the higher order term is defined as constant $d$, and the parameters used are

$$\int_0^1 dx(\frac{x}{2}\ln(\frac{1+x}{1-x}))^2 = 0.6075, \quad \int_0^1 dx(\frac{x^2}{2}\ln(\frac{1+x}{1-x}))^2 = 0.4978 \text{ and}$$

$$\frac{\Delta}{2T}\int_1^{2T/\Delta} dx(\frac{x}{2}\ln(\frac{1+x}{x-1}))^2 - (\frac{\Delta}{2T})^3\int_1^{2T/\Delta} dx(\frac{x^2}{2}\ln(\frac{1+x}{x-1}))^2 \approx 0.6690 - 0.0351\frac{\Delta}{2T}$$

and for $\Delta > 2T$,

$$\frac{(T_1T)^{-1}}{(T_1T)^{-1}\big|_{T=T_c}} = \frac{\Delta}{2T}\int_0^{2T/\Delta} dx(\frac{x}{2}\ln(\frac{1+x}{1-x}))^2 - (\frac{\Delta}{2T})^3\int_1^{2T/\Delta} dx(\frac{x^2}{2}\ln(\frac{1+x}{1-x}))^2 + e(\frac{2T}{\Delta})^\beta + f$$

$$= 0.0571(\frac{2T}{\Delta})^4 + 0.0212(\frac{2T}{\Delta})^6 + e(\frac{2T}{\Delta(T)})^\beta + f$$

Here the higher order terms are defined as $e(\frac{2T}{\Delta(T)})^\beta + f$ where e, f, $\beta$ are constant.

The numerical calculations are

$$\frac{\Delta}{2T}\int_0^{2T/\Delta} dx(\frac{x}{2}\ln(\frac{1+x}{1-x}))^2 \approx \frac{(2T/\Delta)^4}{5} + \frac{2(2T/\Delta)^6}{21}$$

$$(\frac{\Delta}{2T})^3\int_0^{2T/\Delta} dx(\frac{x^2}{2}\ln(\frac{1+x}{1-x}))^2 \approx \frac{(2T/\Delta)^4}{7} + \frac{2(2T/\Delta)^6}{27}$$

After taking note of the conditions, we get

$$\frac{(T_1T)^{-1}}{(T_1T)^{-1}\big|_{T=T_c}} = 1 + 0.6426\frac{\Delta(T)}{2T} - 0.4978(\frac{\Delta(T)}{2T})^3 \quad \text{,for } \Delta(T) < 2T$$

$$= 0.0571(\frac{2T}{\Delta(T)})^4 + 0.0212(\frac{2T}{\Delta(T)})^6 + 1.0665(\frac{2T}{\Delta(T)})^{0.46} \quad \text{,for } \Delta(T) > 2T$$

(13)

Eq.(13) is the simplified equation of the ratio of the nuclear spin lattice relaxation rate of superconducting state to normal state for $^3$He A-phase superfluidity. The numerical results are shown in Figure.2 . It is clear that there is the coherence peak near $T_c$ in $^3$He A-phase superfluidity.

Case III  p-wave superconductors with 3D-linenode

The p-wave superconductors are one of the interesting superconductors. There are linenodes in the order parameters. Here the order parameter of p-wave superconductor is $\Delta_k^p(T) = \Delta^p(T)\cos\theta$ in 3 - dimensional system. In this case, we can get $f(k) \equiv f(\theta) = \cos\theta$. The density of state can be calculated from Eq.(5), we get [7]

$$N(x) = \frac{\pi}{2}x, \qquad x \leq 1$$
$$= x\sin^{-1}(1/x), \qquad x \geq 1 \qquad (14)$$

Here $x = \frac{\varepsilon}{\Delta}$. Substitution Eq.(14) into Eq.(1), we find

$$\frac{(T_1T)^{-1}}{(T_1T)^{-1}\big|_{T=T_c}} = \frac{\Delta}{2T}\int_0^1 dx(\frac{\pi}{2}x)^2 \operatorname{sech}^2(\frac{\Delta x}{2T}) + \frac{\Delta}{2T}\int_1^{2T/\Delta} dx(x\sin^{-1}(\frac{1}{x}))^2 \operatorname{sech}^2(\frac{\Delta x}{2T})$$

Within the same process, we get for $\Delta(T) < 2T$,

$$\frac{(T_1T)^{-1}}{(T_1T)^{-1}\big|_{T=T_c}} = \frac{\Delta}{2T}\int_0^1 dx(\frac{\pi}{2}x)^2(1-(\frac{\Delta x}{2T})^2) + \frac{\Delta}{2T}\int_1^{2T/\Delta} dx(x\sin^{-1}(\frac{1}{x}))^2(1-(\frac{\Delta x}{2T})^2) + h$$

$$= (0.6680 + h) + 0.2085\frac{\Delta}{2T} - 0.4935(\frac{\Delta}{2T})^3$$

Here the higher order term is constant h, and the parameters used are

$$\int_0^1 dx(\frac{\pi}{2}x)^2 = 0.8225, \quad \int_0^1 dx(\frac{\pi}{2}x^2)^2 = 0.4935 \text{ and}$$

$$\frac{\Delta}{2T}\int_1^{2T/\Delta} dx(x\sin^{-1}(\frac{1}{x}))^2 - (\frac{\Delta}{2T})^3 \int_1^{2T/\Delta} dx(x^2\sin^{-1}(\frac{1}{x}))^2 \approx 0.6680 - 0.6140\frac{\Delta}{2T}.$$

And for $\Delta(T) > 2T$,

$$\frac{(T_1T)^{-1}}{(T_1T)^{-1}\big|_{T=T_c}} = \frac{\Delta}{2T}\int_0^{2T/\Delta} dx(\frac{\pi}{2}x)^2 - (\frac{\Delta}{2T})^3 \int_1^{2T/\Delta} dx(\frac{\pi}{2}x^2)^2 + j(\frac{2T}{\Delta})^\gamma + k$$

$$= 0.3290(\frac{2T}{\Delta})^2 + j(\frac{2T}{\Delta(T)})^\gamma + k$$

Here the higher order terms are $j(\frac{2T}{\Delta(T)})^\gamma + k$ where $j, k, \gamma$ are constant. After taking conditions, we get

$$\frac{(T_1T)^{-1}}{(T_1T)^{-1}\big|_{T=T_c}} = 1 + 0.2805\frac{\Delta(T)}{2T} - 0.4935(\frac{\Delta(T)}{2T})^3 \qquad \text{,for } \Delta(T) < 2T$$

$$= 0.3860(\frac{2T}{\Delta(T)})^{1.59} + 0.3290(\frac{2T}{\Delta(T)})^2 \qquad \text{,for } \Delta(T) > 2T \qquad (15)$$

Eq.(15) is the simplified equation of the ratio of the nuclear spin lattice relaxation rate of superconducting state to normal state for p-wave superconductors. The numerical results are shown in Figure.3 .It is clear that there is the coherence peak near c T in p-wave superconductors with linenodes order parameter.

### 3. Result and Discussion

The simple expressions for the ratio of normal state to superconducting state of spin lattice relaxation rate of unconventional superconductors are derived from the BCS weak-coupling equation. We adopted the main ideal of Parker and Haas[7] that used the approximation of density of state near $T_c$ as $N(E) = F(E) + 1$ and calculated the effect of $F(E)$ on the spin lattice relaxation rate . In our calculation, we use the approximation of "$\tanh x$". And to do the calculations more accurately, we keep the higher order terms as some parameters that they should be zero if we use $\tanh x \cong 1$ for x >1. The higher order terms are introduced as a the trial function that can give the results agree with the continuity condition of $\dfrac{(T_1 T)^{-1}}{(T_1 T)^{-1}\big|_{T=T_c}}$. Finally, the $\dfrac{(T_1 T)^{-1}}{(T_1 T)^{-1}\big|_{T=T_c}}$ equations of d-wave, $^3$He A-phase and p-wave superconductors are shown as Eq.(10), Eq.(13), and Eq.(15) respectively. All of them are in the power series of $\Delta(T)$ and temperature. As low temperature, the $\dfrac{(T_1 T)^{-1}}{(T_1 T)^{-1}\big|_{T=T_c}}$ is in the power law of T that agree with the work of ref.[6] .And as high temperature, the $\dfrac{(T_1 T)^{-1}}{(T_1 T)^{-1}\big|_{T=T_c}}$ is in the power law of $\Delta(T)$ that agree with the work of ref.[7]. For numerical calculations, the $\Delta(T)$ with determined parameters $\Delta(0)$, $T_c$ and $\omega_D$ can be found by using the BCS gap, Eq.(3) and Eq.(4). Because $\dfrac{(T_1 T)^{-1}}{(T_1 T)^{-1}\big|_{T=T_c}}$ are in the power series of the ratio of $\Delta(T)$ to T , we can find this ratio from Eqs.(3-4) . The ratio of $\Delta(T)$ to T can be found by using Eq.(3) only which do not depent on $\Delta(0)$. However, we must calculate the $\Delta(0)$ and $T_c$ from Eq.(4) to ensure that they are the BCS superconductors. Our results are shown in Figure.1-3. All of figures are shown $\dfrac{(T_1 T)^{-1}}{(T_1 T)^{-1}\big|_{T=T_c}}$ versus temperature with varying the magnitude of order parameters at fix $\omega_D$. Our results clearly show that there is the coherence peak near $T_c$ in all type of order parameters considered.

These results do not agree with the belief that the coherence peak is the only property of s-wave superconductors. We think that these phenomena occurred because these systems have the same mechanism as s-wave superconductors ,Cooper pair, so the measurement of the nuclear spin-lattice relaxation should also be the same as s-wave superconductors. So the question still remain that why the experimentalists can not detect coherence peak in these superconductors before. Because almost of d-wave and p-wave superconductors are impurities superconductors, we think that the impurities may be the important factors that destroyed the coherence peak below $T_c$ in these materials.

**4.Conclusion**

The simple expression for the ratio of normal state to superconducting state of spin lattice relaxation rate of unconventional superconductors are shown. The $\dfrac{(T_1 T)^{-1}}{(T_1 T)^{-1}\big|_{T=T_c}}$ are given in the power series of T and $\Delta(T)$ in low and high temperature respectively. However, the unconventional superconductors of our consideration can show the coherence peak below $T_c$. These results do not agree with the belief that the coherence peak is the only property of s-wave superconductors.

**Acknowledgement**

The author would like to thank Professor Dr.Suthat Yoksan for the useful discussion and also thank Thai Research Fund and Office of Higher Education Commission, Faculty of Science of Srinakharinwirot University, and ThEP Center for the financial support.

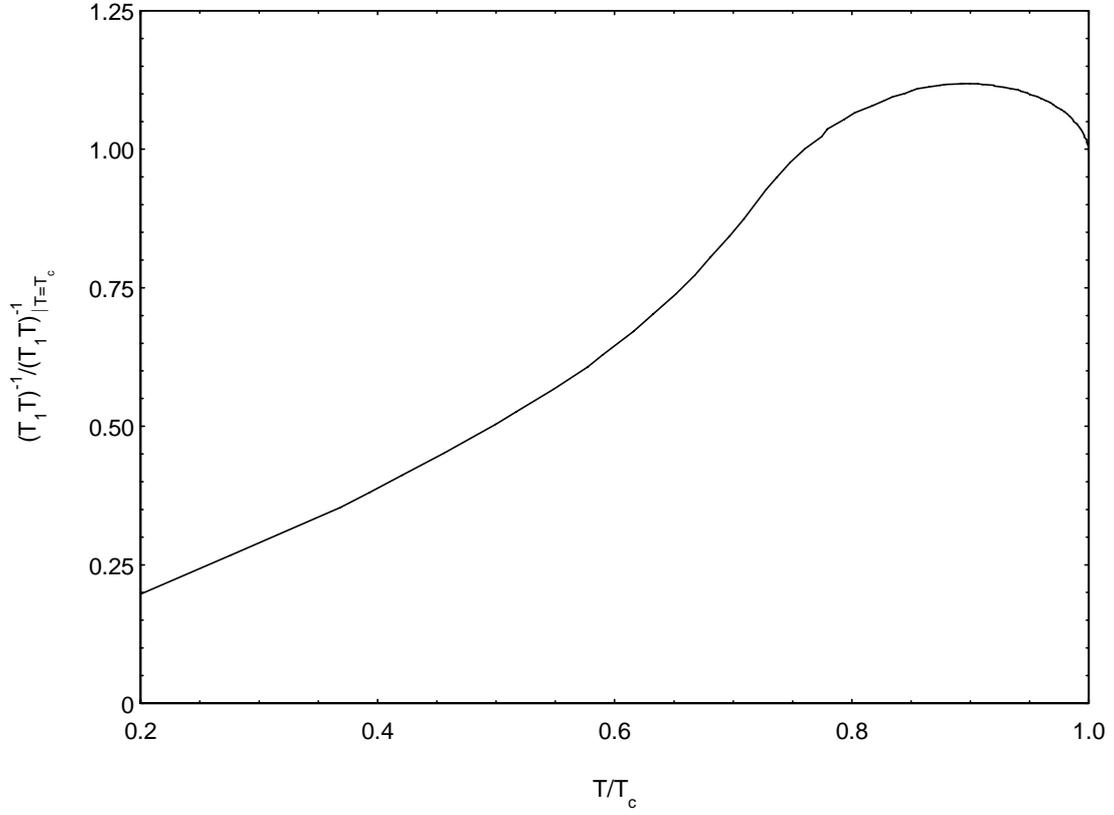

Figure 1. Shown $\dfrac{(T_1T)^{-1}}{(T_1T)^{-1}\big|_{T=T_c}}$ of $d_{x^2-y^2}$-wave superconductors. The parameters used are $\omega_D$ = 700 K and with varying $\Delta(0)$ =190K,150K,100K that give $T_c$ = 100.2 K,78.8K ,52.6 K respectively. These results are coincided within numerical accuracy.

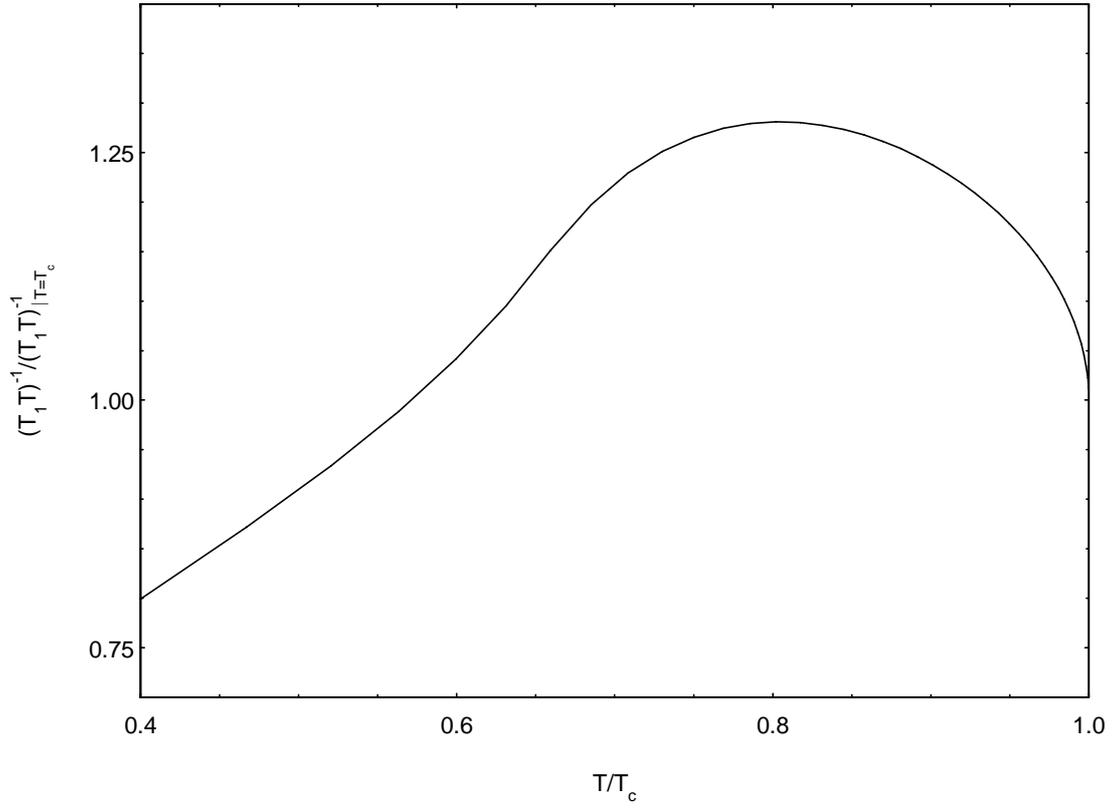

Figure 2. Shown $\dfrac{(T_1 T)^{-1}}{(T_1 T)^{-1}\big|_{T=T_c}}$ of $^3He$ A-phase. The parameters used are $\omega_D$ = 10 mK and with varying $\Delta(0)$ =4.7 mK, 4.4 mK, 4.1 mK that give $T_c$ = 3.03 mK, 2.82 mK, 2.62 mK respectively. These results are coincided within numerical accuracy.

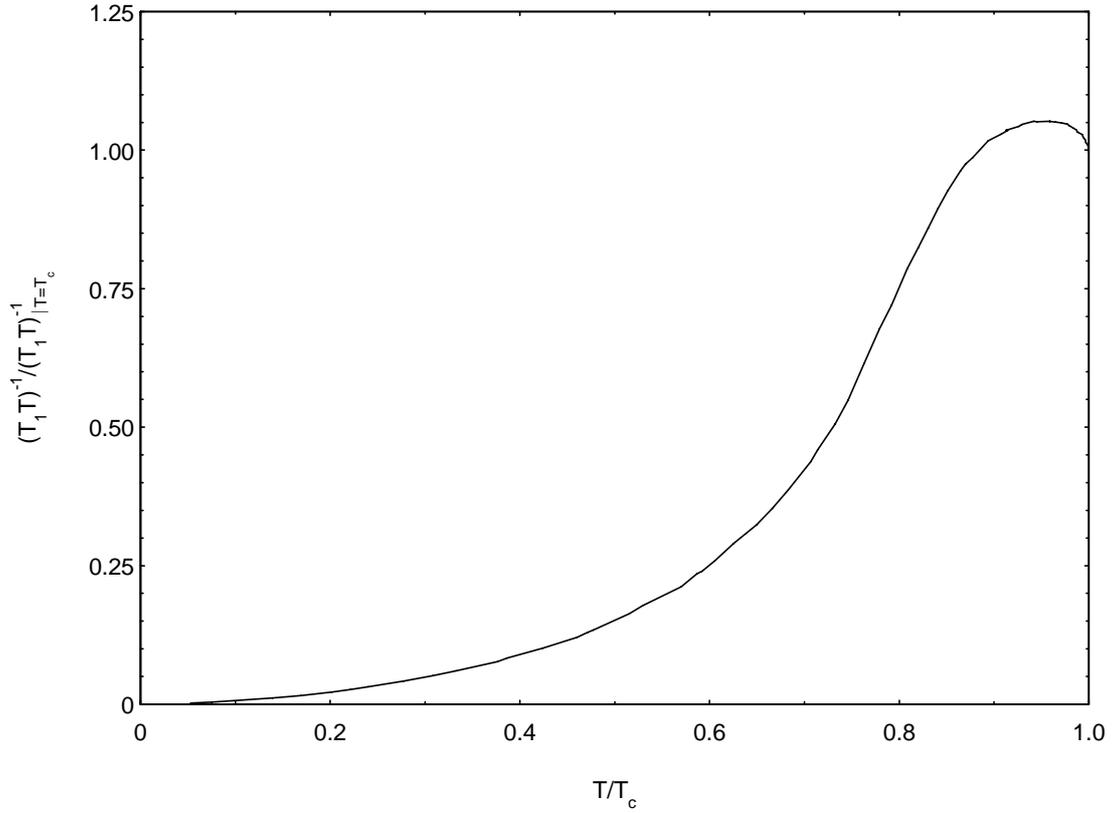

Figure 3. Shown $\dfrac{(T_1T)^{-1}}{(T_1T)^{-1}\big|_{T=T_c}}$ of p-wave superconductors with 3D linenodes. The parameters used are $\omega_D$ = 5 K and with varying $\Delta(0)$ = 1.7 K, 1.6 K, 1.5K that give $T_c$ = 0.53 K, 0.50 K, 0.45 K respectively. These results are coincided within numerical accuracy.